\documentclass{article}
\usepackage[utf8]{inputenc} 
\usepackage[T1]{fontenc}    

\usepackage{arxiv}
\usepackage{amsmath,amsfonts}
\usepackage{hyperref}       
\usepackage{url}            
\usepackage{booktabs}       
\usepackage{amsfonts}       
\usepackage{nicefrac}       
\usepackage{microtype}      
\usepackage{lipsum}
\usepackage{graphicx}
\usepackage{array}
\usepackage[caption=false,font=normalsize,labelfont=sf,textfont=sf]{subfig}
\usepackage{textcomp}
\usepackage{stfloats}
\usepackage{url}
\usepackage{verbatim}
\usepackage{siunitx}
\usepackage{tabularx}
\usepackage{float}
\title{Simulation of silicon ridge waveguide enhanced two-photon absorption from femtosecond pulses}

\author{
 Cael Warner \\ 
  Electrical and Computer Engineering Department \\
  University of Alberta \\
  Edmonton, AB T6G 2R3 \\
  \texttt{spencerw@ualberta.ca} \\
  \And
 Ruoheng Zhang \\ 
  Electrical and Computer Engineering Department \\
  University of Alberta \\
  Edmonton, AB T6G 2R3 \\
  \texttt{ruoheng@ualberta.ca}
}

\begin{document}
\maketitle
\begin{abstract}
Field enhancement of two-photon absorption from a 50-\si{fs} pulse on a silicon ridge waveguide is simulated for a varying energy downward propagating $800\pm9.42$ \si{nm} wavelength plane-wave  orthonormal to a waveguide in 2D FDTD using ANSYS Lumerical FDTD. Energy absorbed by the waveguide is enhanced due to mode confinement within the standard 500 \si{nm} width, 130 \si{nm} height silicon ridge waveguide on 90 \si{nm} thick silicon and 3 \si{\micro m} silicon dioxide insulator.
\end{abstract}


\section{Introduction}
Silicon photonic devices are finding rapid application as essential components in integrated electronics and data centers \cite{li2015silicon}. However, process variation in fabrication results in variation of passive device performance in terms of effective refractive index, $n_{eff}$. Production of multiple waveguides or use of micro-heaters, to modify the material for desired performance, is a wasteful process leading to significant energy usage and multi-billion dollar losses for data-center operation. Alternatively, free carriers can be excited via two-photon absorption (TPA) in silicon thereby causing defects in the crystal lattice which modify $n_{eff}$ to achieve desired device performance. Changes in $n_{eff}$ can be measured directly using planar thin-film silicon microring resonators on insulating silicate substrates. Quasi-neutral non-thermal melting retains geometric structure of the device but modifies the crystal lattice microstructure, and the device's optical properties \cite{zhang2024precision}. Non-thermal melting is attained near the band-gap energy of silicon, $E_\mathrm{g}=1.12$ \si{eV}, using $\lambda_0=800$ \si{nm} wavelength laser ablation to excite TPA. At an orthonormal angle of incidence with respect to the waveguide axis, $\lambda_0=800$ \si{nm} light couples strongly to the standard ridge waveguide dimensions (500 \si{nm} width and 130 \si{nm} height), such that the intensity and TPA of a 50-\si{fs} 300 \si{nJ} pulse is enhanced. Therefore, with a low repetition rate, only the ridge wave-guide's microstructure defects and $n_{eff}$ are modified with negligible thermal melting or material strain \cite{zhang2024precision}. This is in contrast to ordinary thermal ablation, where above threshold melting fluence causes craters in silicon \cite{bonse2002femtosecond}. Few studies simulate the initial two-dimensional field enhancement from two-photon absorption \cite{mukherjee2012simulation}, which occurs instantaneously compared with the free carrier absorption (over 6 \si{ps}) \cite{rousse2001non}. Control of the interference mechanism may allow tailored in-situ micro-structural modification of silicon chips in future re-purposing or resonator tuning, enabling desired device performance without contributing additional electronic waste by fabricating multiple passive devices.

\section{Two-photon absorption Model}
\label{sec:headings}
General nonlinear polarizability 
\begin{equation}\label{eq:1}
    \mathbf{P}^{(n)}(z,\omega)=D\epsilon_0\mathbf{\chi}^{(n)}(\omega;\omega_1,\omega_2,...,\omega_n)\mathbf{E}(z,\omega_1)\mathbf{E}(z,\omega_2)...\mathbf{E}(z,\omega_n),
\end{equation}
Exhibits a degeneracy factor of $D=n!/m!=3$ for Raman absorption, compared with $D=n!/m!=6$ for TPA, where $n$ represents the order of the polarizability and $m$ represents the number of waves. If we assume frequency-domain two-wave degeneracy for TPA by an isotropic material, then the two-photon imaginary third-order susceptibility, $\chi"^{(3)}_\mathrm{TPA}$, is numerically equivalent to the Raman imaginary third-order susceptibility, $\chi"^{(3)}_\mathrm{R}$, apart from a constant factor of 2 from $D$. ANSYS Lumerical\textsuperscript{\textregistered} FDTD uses a built-in Chi3 Raman Kerr material plugin model based on earlier work by Blow \& Wood \cite{blow1989theoretical},and Goorjian \& Taflove \cite{goorjian1992direct}. In simplified terms, the ANSYS Lumerical model represents \cite{ansysOptics}
\begin{equation}\label{eq:2}
    \mathbf{P}(t)=\epsilon_0\chi^{(1)}\mathbf{E}(t)+\epsilon_0\eta\chi_0^{(3)}\mathbf{E}^3(t)+\epsilon_0(1-\eta)\mathbf{E}(t)(\chi^{(3)}_R(t)\ast\mathbf{E}^2(t)),
\end{equation}
Where the susceptibility has a Lorentzian spectrum \cite{goorjian1992direct}
\begin{equation}\label{eq:3}
    \chi_R^{(3)}(\omega)=\frac{\chi_0^{(3)}\omega_R^2}{\omega_R^2-2j\omega\delta_R-\omega^2},
\end{equation}
Such that the time-domain response of $\chi_R^{(3)}(t)$ is solved from a Fourier transform. We note that this form is from the ANSYS Lumerical documentation, and has some numerical coefficients which differ from Goorjian \& Taflove's \cite{goorjian1992direct}. With modification by a factor 2, $\chi"^{(3)}_\mathrm{R}=2\chi"^{(3)}_\mathrm{TPA}$, the Chi3 Raman Kerr material plugin can represent measured TPA coefficient $\beta$ \cite{bristow2007two}, 
\begin{equation}\label{eq:4}
    \chi_0^{(3)}(\omega)=\frac{\epsilon_0 cn_0^2\beta[(\omega_R^2-\omega^2)^2+4\omega^2\delta_R^2]}{24\omega^2\omega_R^2\delta_R},
\end{equation}
Where $\omega_R$ is the Raman frequency, and $\delta_R$ is the Raman spectrum full width at half maximum (FWHM), also experimentally measured \cite{aggarwal2011measurement}. Provided measurements of both Kerr nonlinear refractive index, $n_2$, and $\beta$, specific tuning of a dimensionless coefficient, $\eta$, represents proportion of real Raman and Kerr susceptibility added at a single wavelength  such that
\begin{align}
    \chi'^{(3)} & = (1-\eta)\chi_0^{(3)}+\eta\chi'^{(3)}_R \tag{5a}\label{eq:5a}\\
        \eta &=\frac{4\left\{\delta_R\left[(n_2\omega_R^2/c\beta - 4\delta_R)\omega^2\right]-(\omega_R^2-\omega^2)^2\right\}}{\omega^2\left\{|\omega_R^2-\omega^2|+4\delta_R^2\right\}},\tag{5b}\label{eq:5b}
\end{align}
Where $\eta\in(0,1)$. Alternatively, given an impulse response for the Raman contribution, 
\begin{equation}
  g_R(t)=\left(\frac{1}{\tau_1^2}+\frac{1}{\tau_2^2}\right)\tau_1\exp{\left(-\frac{t}{\tau_2}\right)}\sin{\left(\frac{t}{\tau_1}\right)},\tag{6}\label{eq:6}
\end{equation}
Where $\tau_1=1/\omega_R$ and $\tau_2=1/\delta_R$, one may also solve for $\eta$ as
\begin{equation}
    \eta=\frac{cn(\omega_0)H_R(\Delta \omega)}{\omega_0\chi^{(3)}\mathrm{Im}\left(G_R(\Delta \omega)\right)},\tag{7}\label{eq:7}
\end{equation}
Where $H_R(\Delta \omega)$ is the Raman-shifted spectrum \cite{agrawal2000nonlinear}. For more details of the derivation, please refer to Supplementary material.

Intensity dependent two-photon absorption occurs instantaneously compared with free carrier absorption at room temperature (6 \si{ps}) \cite{korfiatis2007conditions}. Therefore, we initially neglect free carrier absorption for the first single 50-\si{fs} pulse (assuming it is a factor 100 lower frequency), and examine field enhancement from two-photon absorption alone.  The analytical one-dimensional intensity is 
\begin{equation}\tag{8}\label{eq:8}
    I= I_0 \frac{\exp(-\alpha z)}{(\alpha+\beta I_{0})-\beta I_{0}\exp(-\alpha z)},
\end{equation}
Where $\beta$ is the two-photon absorption coefficient and $\alpha$ is the linear absorption coefficient. Absorbed energy density is described by $u_{\mathrm{ABS}}=\int\beta I^2\mathrm{d}t$. Each analytical field profile in one dimension is compared with the two-photon absorption profile in bulk silicon to verify the simulation accuracy of the material plugin model.

\section{Two-photon absorption material model}
ANSYS Lumerical FDTD is used to model two-photon absorption in isotropic silicon with the Chi3 Raman Kerr material plugin. The Kerr refractive index and two-photon absorption coefficient \cite{bristow2007two} are fitted to the Raman spectrum for a linewidth typical of the femtosecond laser. This is achieved by equating the real and imaginary third-order susceptibility of the model to real and imaginary terms measured by Bristow, accounting for a degeneracy factor. Given this derivation (see the Appendix), the corresponding parameter of interest is  $\chi_0^{(3)}(\omega)$ from (\ref{eq:4}), where $\beta$, $n_2$, $\omega_R$, and $\delta_R$ are used from previous measurements \cite{bristow2007two,aggarwal2011measurement}. Table \ref{tab:1} includes a list of these parameters. The imparted intensity on the silicon is varied, $I_0\in(113.2,11320)$ \si{GW/cm^2}, to represent different pulse energies, $U\in(80.56,4319.74)$ \si{nJ}, although the temporal length of the pulse ($\tau=50$ \si{fs}) is the same for each case over a $w(0)=13.4$ \si{\micro m} spot size. The base material of the Chi3 Raman Kerr material plugin is Si (Palik) \cite{palik1998handbook}, which is fitted using a simplex algorithm over the wavelength range $\lambda_0\in(750,850)$ \si{nm}.

\begin{table}[t]
    \scriptsize
    \centering
    \caption{Model properties for Silicon}
    \begin{tabularx}{\textwidth}{|l|l|l|X|}
    \hline
         \textbf{Parameter name} & \textbf{Symbol} & $\lambda_0=800$ \si{nm} & Unit \\\hline\hline
         Linear absorption coefficient & $\alpha $ & 1000 & \si{cm^{-1}} \\\hline
         Two photon absorption coefficient & $\beta$ & 2.000 &\si{cm GW^{-1}} \\\hline
         Raman frequency & $\omega_R$ & 15.56 &\si{THz}  \\\hline
         Raman spectrum FWHM & $\delta\omega_R$ & 2.91 &\si{THz}\\\hline
         Susceptibility model parameter & $\chi^{(3)}_0$ & $1.4361(10^{-12})$ &\si{ cm^{2}V^{-2}}\\\hline
         Kerr contribution & $\eta$ & $1.1607(10^{-3})$ & - \\ \hline
    \end{tabularx}
    \label{tab:1}
\end{table}

\begin{figure}[H]
    \centering
    \includegraphics[width=0.6\linewidth]{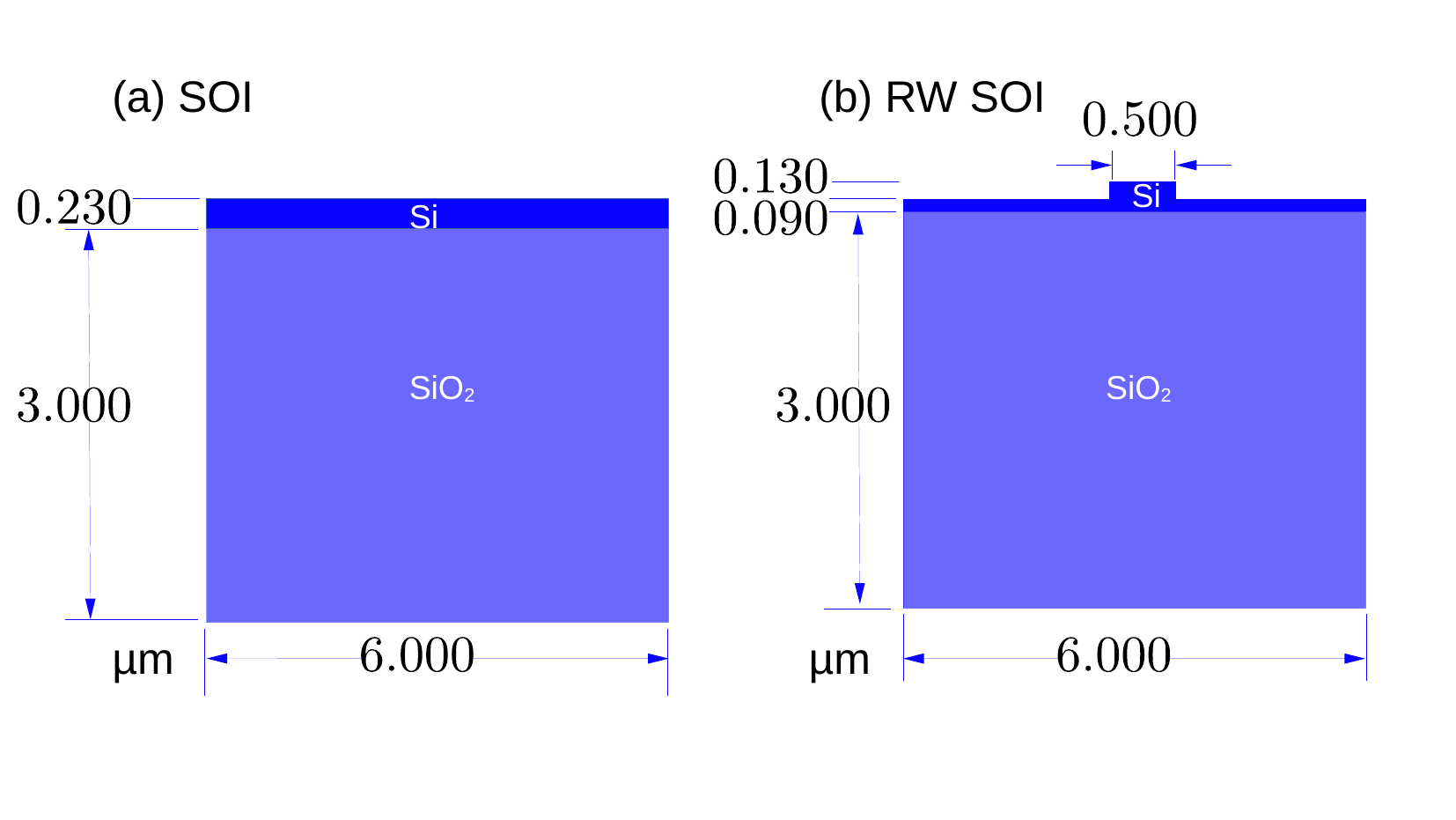}
    \\\includegraphics[width=\linewidth]{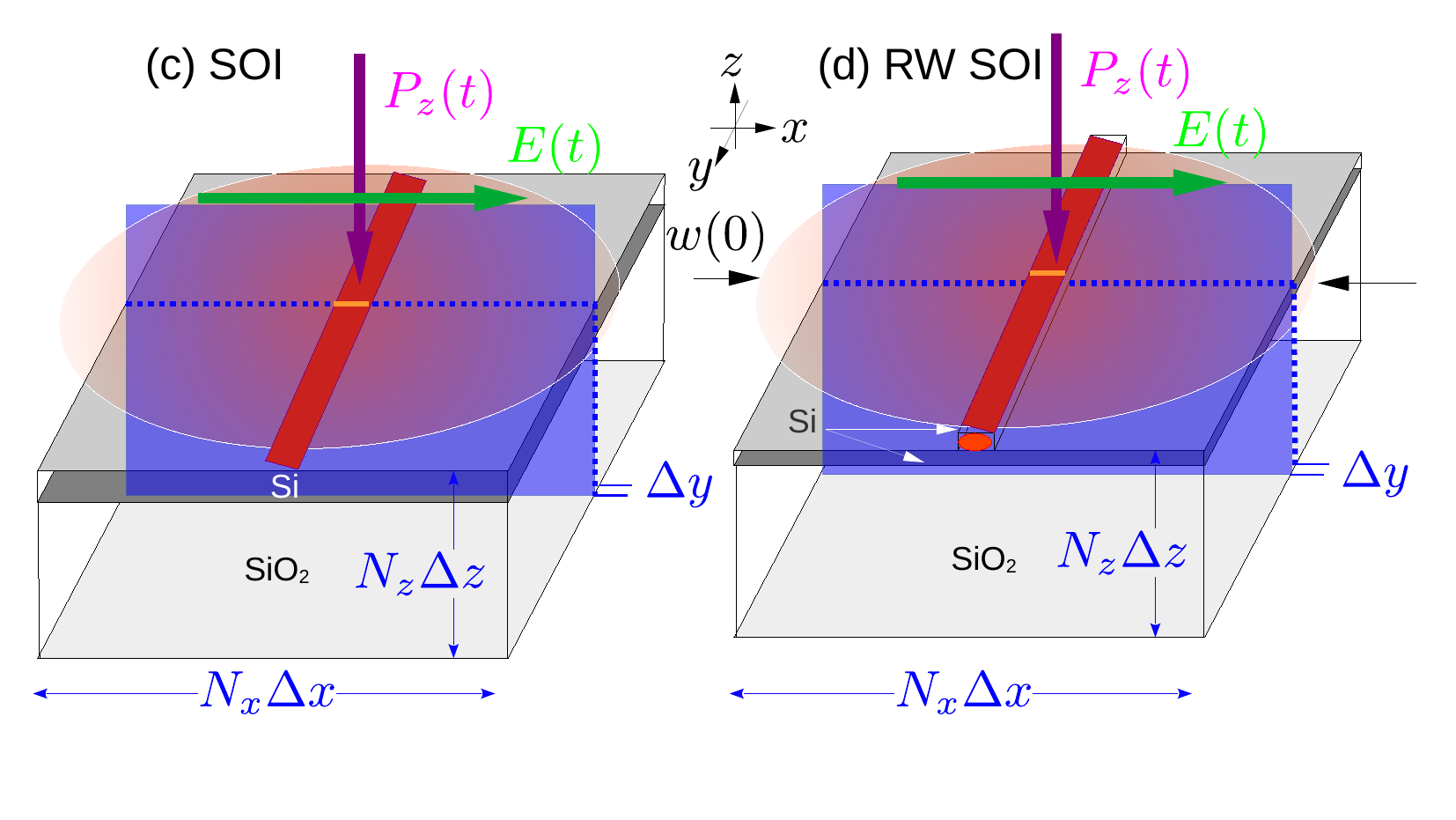}
    \caption{(a) SOI cross-section, (b) RW SOI cross-section, (c) SOI system, and (d) RW SOI system. In (a) and (b) the units are in \si{\micro m}. Each cross section in (a) and (b) is illustrated in (c) and (d), respectively. In (c) and (d), the red spot represents the laser, $\mathrm{P_z}(t)$ is the Poynting vector, and $\mathrm{E}(t)$ is the electric field. Number of spatial units in $x$, $y$ and $z$ Cartesian directions are $N_x$, $N_y$, and $N_z$. At the surface, the beam diameter is $\omega(0)$ and the beam is weakly divergent and approximated as a plane wave to satisfy the periodic boundaries.}
    \label{fig:1}
\end{figure}

\section{Structure and simulation}
\par Fig. \ref{fig:1} illustrates the cross-section geometry and analyzed system for the silicon-on-insulator (SOI) and rectangular ridge-waveguide on insulator (RW SOI) structures.  This structure is compared with the case of bulk silicon, and a silicon on insulator (SOI) structure consisting of a 230 \si{nm} thick silicon layer on top of a 3 \si{\micro m} silicon dioxide layer and an identical thickness silicon substrate. The ridge waveguide silicon on insulator (RW SOI) structure consists of a 500 \si{nm} width, 130 \si{nm} deep silicon ridge waveguide on a 90 \si{nm} thick silicon substrate, placed on top of a 3 \si{\micro m} silicon dioxide layer and arbitrary thickness silicon substrate that extends beyond the FDTD domain. Two-dimensional (2D) simulation in ANSYS\textsuperscript{\textregistered} Lumerical FDTD is made possible by the symmetry of the linearly-polarized plane wave with respect to the structure. Otherwise, three-dimensional (3D) simulation exceeds ordinary work-station memory (\( \sim\) 64 \si{GB}) without providing additional information. 

\par The source condition is a plane wave, in order to better conserve momentum at the side edge (west and east) boundary conditions which are periodic. Although the femto-second laser is typically imparted on a micro-ring structure to measure the change in quality factor with $n_{eff}$, the focal spot is imparted on a rectangular prism section of the waveguide with a cross-section represented in the simulation. The plane wave has a pulse-length of 50-\si{fs}, with a Gaussian profile, and a wavelength of $\lambda_0\in(790.89,809.53)$ \si{nm} (were the center wavelength is $\lambda_c=800$ \si{nm}). The time domain is $t_f=200$ \si{fs} with a peak field intensity occurring at $t_c=100$ \si{fs}. Different uniform grids were used in each material region for sufficient numerical accuracy. It is necessary to include a low time stability factor of $\delta t=0.05\Delta t_\mathrm{FS}$, or $5\%$ the optimal free space stability factor, and a grid spacing of 50 units per material wavelength (including linear index of refraction). Although relatively coarse for the nonlinear terms \cite{goorjian1992direct}, the necessary memory does not exceed a single work-station and the convergence error is satisfactory. Additional grid points may be required to describe longer pulses or continuous waves which exhibit filamentous modes within the rectangular waveguide. Time-domain intensity is averaged for one period of the 800 \si{nm} center wavelength, near $t_c+\Delta t$, where $\Delta t$ accounts for the reduced group velocity of the wave in the silicon. Stabilized, stretched-coordinate perfectly matched layer (SSCPMLs) are used on each top and bottom (North and South) boundary. 

Spatial grid spacing for each simulation is uniform in each material region, and consistent, tested to ensure similar solutions are obtained with mesh refinement. In the case of either material, silicon (Si) or silicon dioxide (SiO\textsubscript{2}), $\Delta x=\Delta y=4.32432$ \si{nm}. In the case of  air, the mesh is auto non-uniform with conformal variant 0, a mesh accuracy of 3, and a minimum mesh step of $\Delta x_\mathrm{FS,min}=\Delta y_\mathrm{FS,min}=0.25$ \si{nm} (typical $\Delta x_\mathrm{FS}=4.32432$ \si{nm} and $\Delta y_\mathrm{FS} = 55.6$ \si{nm}).  Intensity and energy absorption profiles are compared with their analytical solutions, and the absorbed energy is plotted with respect to incident fluence.

Figs. \ref{fig:1}a-\ref{fig:1}b illustrate the 2D cross section geometry of the SOI and RW SOI structures and their dimensions in units of \si{\micro m}. Each 2D  cross section is a single (blue) slice of the 3D system represented in Figs. \ref{fig:1}c-\ref{fig:1}d.  Absorbed energy in the 2D simulated material slice with dimensions $N_x\Delta x\times N_z\Delta z$, is added over the area of the laser spot (red spot with diameter given by $w(0)$) for each discrete depth slice $\Delta y$. The laser spot is assumed to have a top-hat profile with uniform intensity over its area, such that each slice has equivalent energy density to satisfy 2D symmetry along $x$ and $z$ in the 3D system.

\section{Time averaged field intensity}
\begin{figure}[H]
    \centering
    \includegraphics[width=0.33\linewidth]{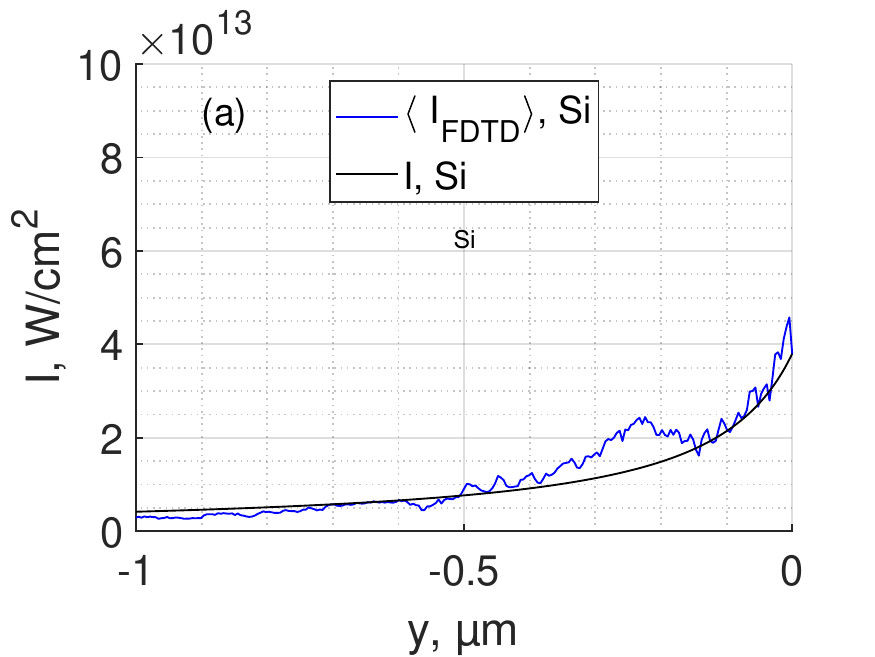}
       \includegraphics[width=0.33\linewidth]{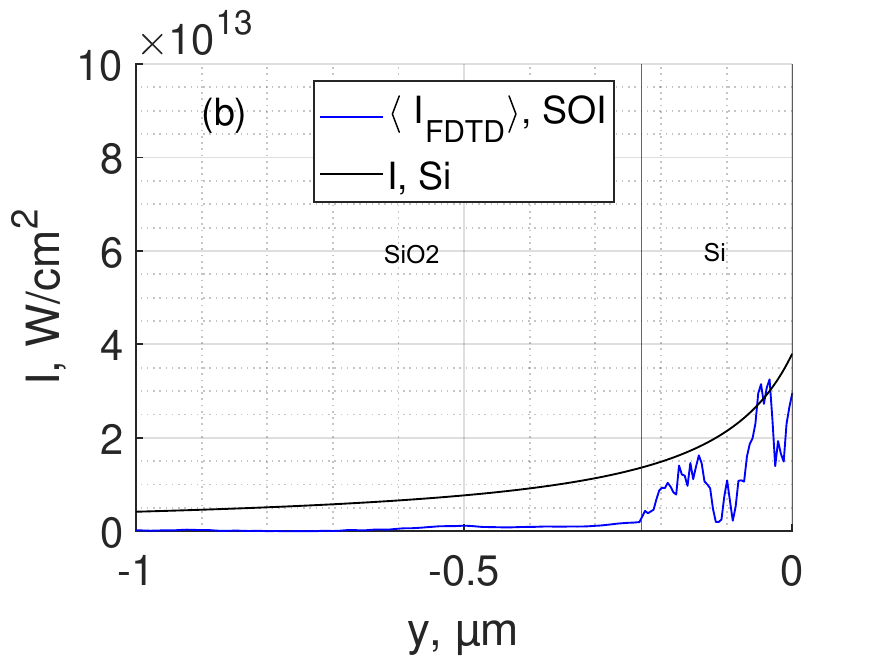}
          \includegraphics[width=0.32\linewidth]{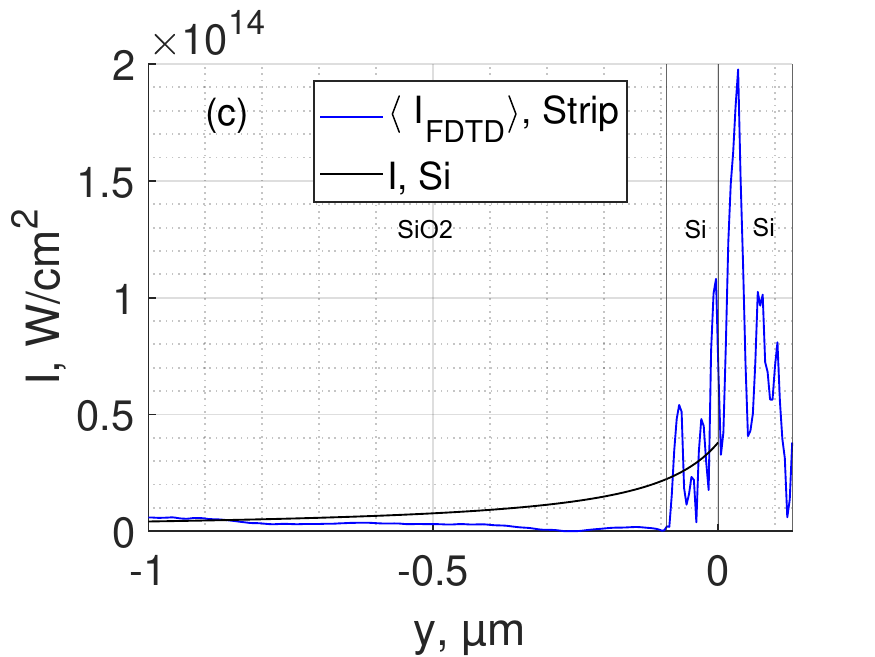}
    \caption{The field intensity plotted for the case of (a) the silicon substrate, (b) silicon on insulator, and (c) ridge waveguide. Note that in (c) the intensity axis is scaled by a factor 2, and the peak intensity is a factor of 4 greater than the surface intensity of silicon. }
    \label{fig:2}
\end{figure}

Given the current measurement error of Kerr refractive index for $\lambda=800$ \si{nm} at high intensity, particularly when compared to the fitted Raman impulse response, two different values of $\eta$ were solved for ($\eta=0.001610736$ and $\eta=0.3575$) that both resemble the two-photon absorption intensity profile but predict different modulation of the absorbed energy density. In order to compensate, we choose a rounded median value of $\eta=0.1$ that was found to best qualitatively represent the two-photon absorbed energy density. In the case of the 67.08 \si{\micro J} pulse on a pure silicon substrate, the fields are attenuated within 200 \si{nm} from the surface, as demonstrated by the absorption profile in intensity plotted in Fig. \ref{fig:2}a.  Although similar field attenuation occurs in the silicon on silicon dioxide, reflections confine the electromagnetic energy to the silicon. Last is the case of the ridge waveguide, where the intensity along the center-line increases toward the center of the waveguide due to electromagnetic mode confinement.  Therefore, interference and diffraction from the structure serve to increase the TPA.

\begin{figure}[H]
    \centering
    \includegraphics[width=0.33\linewidth]{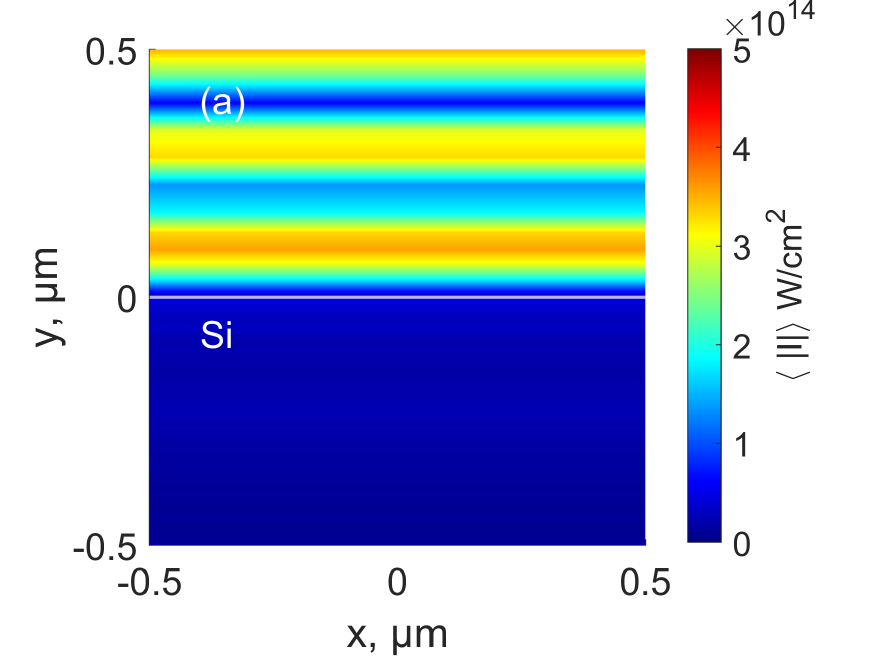}
       \includegraphics[width=0.33\linewidth]{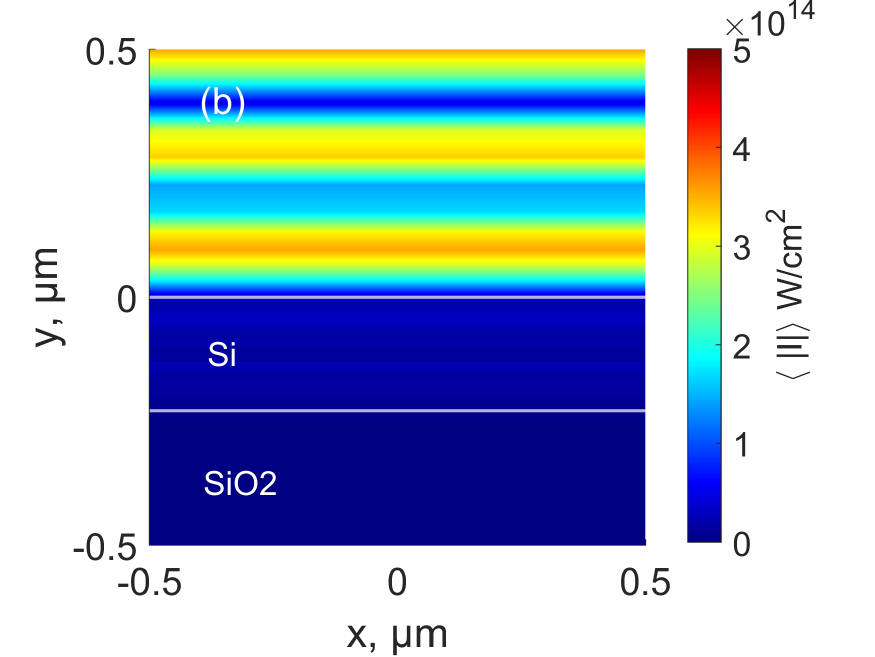}
          \includegraphics[width=0.32\linewidth]{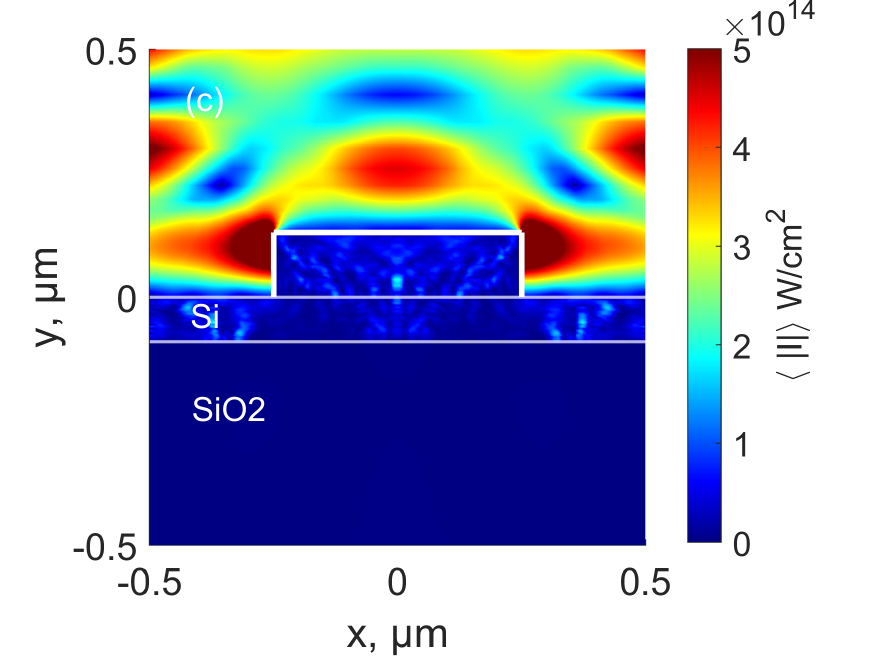}
    \caption{The field intensity plotted for the case of (a) the silicon substrate, (b) silicon on insulator, and (c) ridge waveguide. }
    \label{fig:3}
\end{figure}

The numerically evaluated two-dimensional single cycle time-averaged field intensity is plotted in Fig. \ref{fig:3}. In these plots, the relative field enhancement with respect to the planar structure can be directly observed.

\section{Absorbed energy with respect to fluence}
Absorbed energy in silicon, $U_{ABS}=\int_{\mathbf{x}}\int_t\beta I^2 \mathrm{d}^3\mathbf{x}\mathrm{d}t$, is plotted with respect to the fluence imparted, $F=U/\pi w(0)^2$, on the surface of the slab given a beam-waist of $w(0)=13.4$ \si{\micro m}. Mode confinement in the silicon waveguide enhanced the absorbed energy at lower fluence. 
The absorbed fluence is calculated using a simple sum of the discrete energy distributed throughout a volume the size of the focal spot assumed to have uniform energy density per each discrete two-dimensional slice representing the simulation domain. Symmetry of the two-dimensional plane into the three-dimensional system is made given the linearly polarized beam with a plane-wave approximation. This approximation over-estimates the energy, but helps enforce the two-dimensional symmetry.
Despite discrepancy describing the beam profile, two-photon absorbed energy of a plane wave in bulk silicon follows a parabolic function for low energies with respect to the Fluence, as expected from the theoretical description. In contrast, the function is different for the SOI and RW on SOI structures. Energy absorbed in the waveguide alone, with respect to the incident fluence on the waveguide, demonstrates a different relationship compared with bulk silicon. Increased two-photon absorption is due to mode confinement in the silicon structure, with a weak relation to nonlinear optical properties of silicon. Linear absorption in the silicon is also enhanced by mode confinement.

\begin{figure}[H]
    \centering
    \hspace{-10mm}\includegraphics[width=0.33\linewidth]{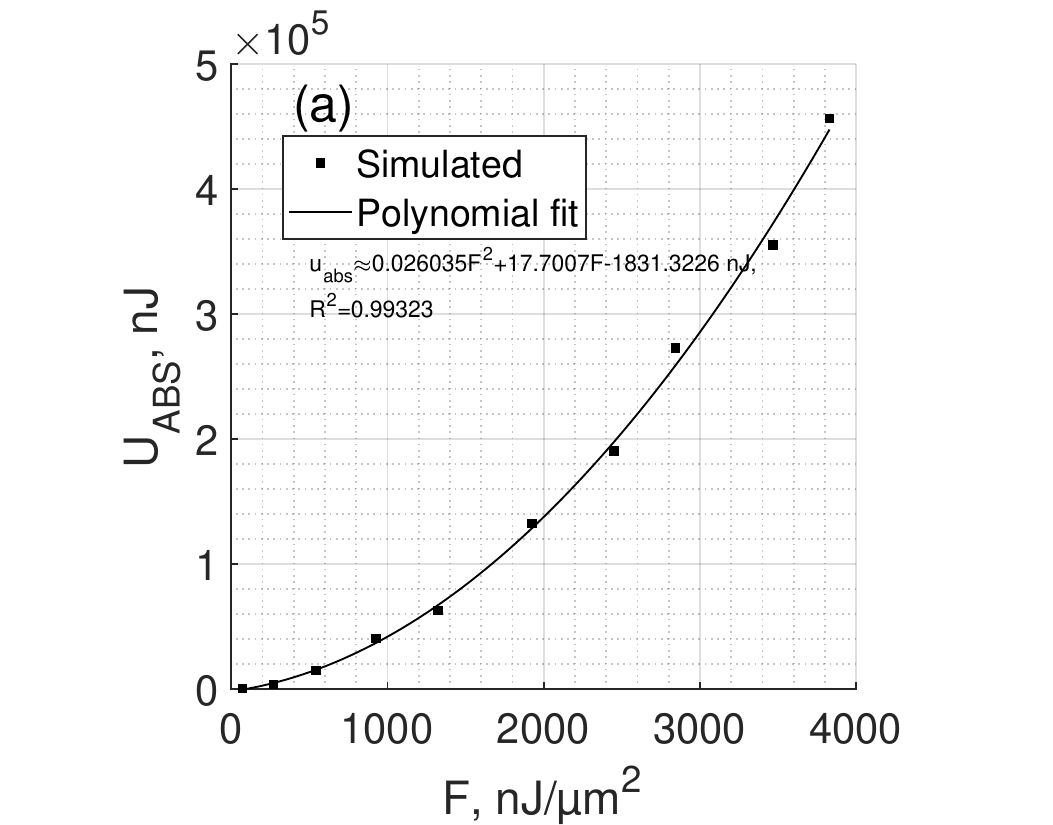}
    \hspace{-10mm}\includegraphics[width=0.33\linewidth]{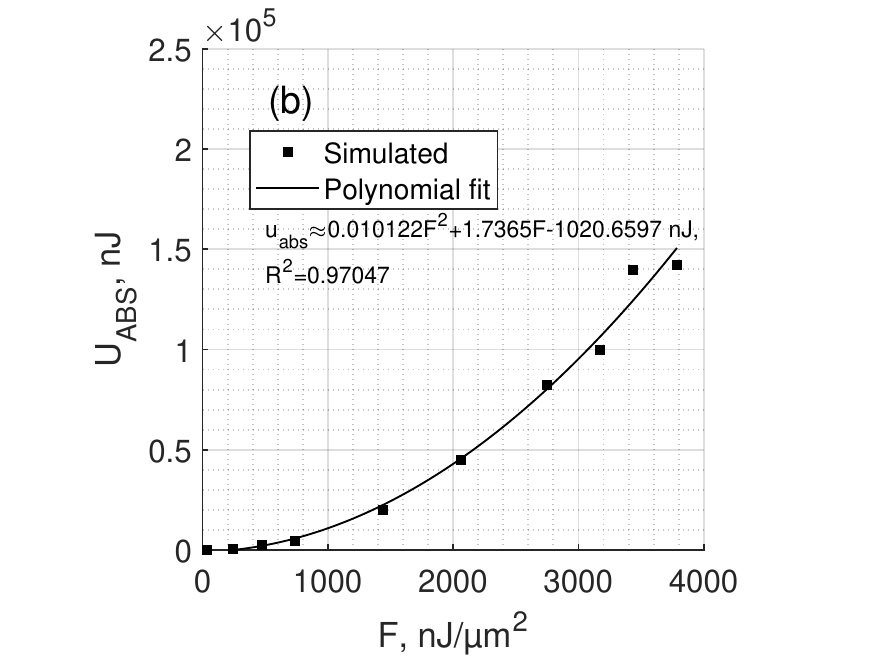} \hspace{-10mm}\includegraphics[width=0.32\linewidth]{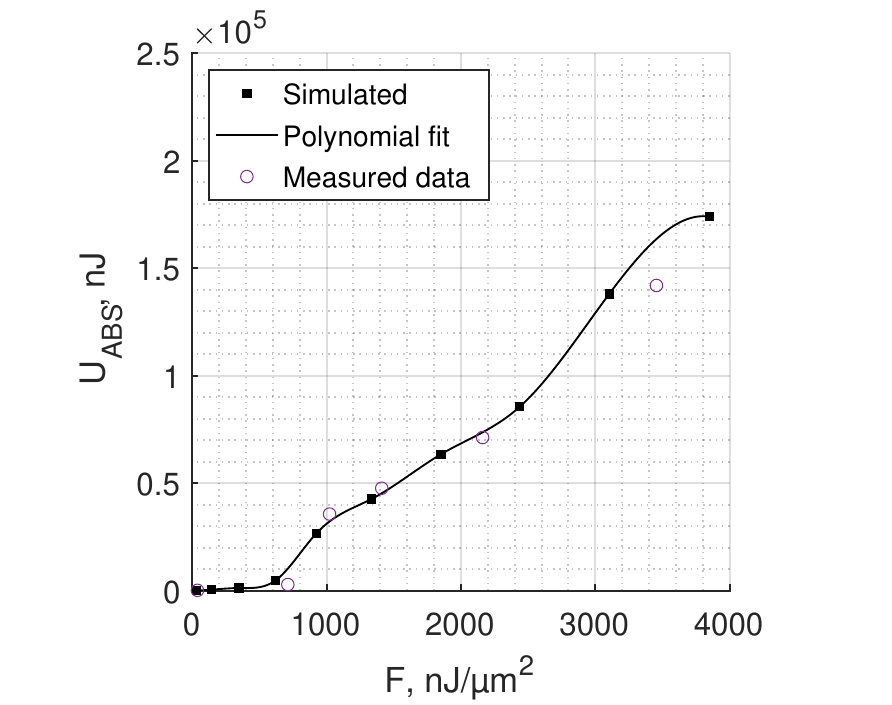}
    \caption{Two-photon-absorbed energy, $\mathrm{U}_\mathrm{ABS}$, in nano-Joules, \si{nJ}, with respect to incident laser fluence, $F$, for a flat-top beam profile with a beam-waist of $w(0)=13.4$ \si{\micro m} in the case of (a) bulk silicon; (b) silicon-on-insulator; and (c) ridge-waveguide on silicon-on-insulator. In (a) the quadratic polynomial fit is highly correlated to simulated data-points (black squares). In silicon on insulator, a similar quadratic polynomial fit is obtained but with less than half the two-photon energy absorption of bulk silicon due to transmission in the silicon dioxide layer. In (c), the quadratic absorption is perturbed given the local field enhancement in the waveguide. Experimental data from Fig. 7.7 in Zhang, 2024 \cite{zhang2024precision} (refer to Supplementary document) is represented by purple circle markers. }
    \label{fig:4}
\end{figure}

\begin{figure}[H]
    \centering
    \hspace{-10mm}\includegraphics[width=0.33\linewidth]{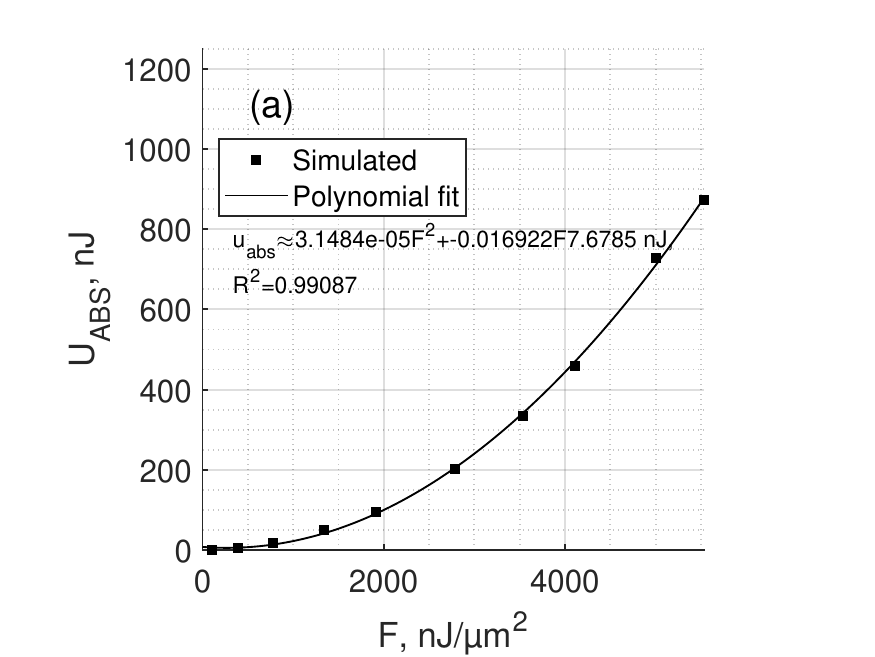}\hspace{-10mm}
    \includegraphics[width=0.33\linewidth]{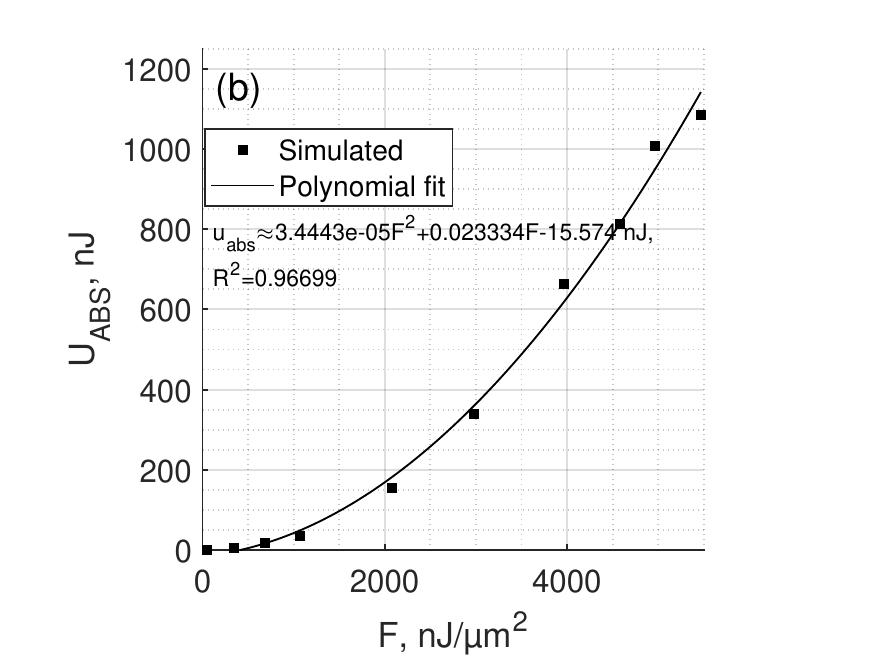}\hspace{-10mm}
    \includegraphics[width=0.32\linewidth]{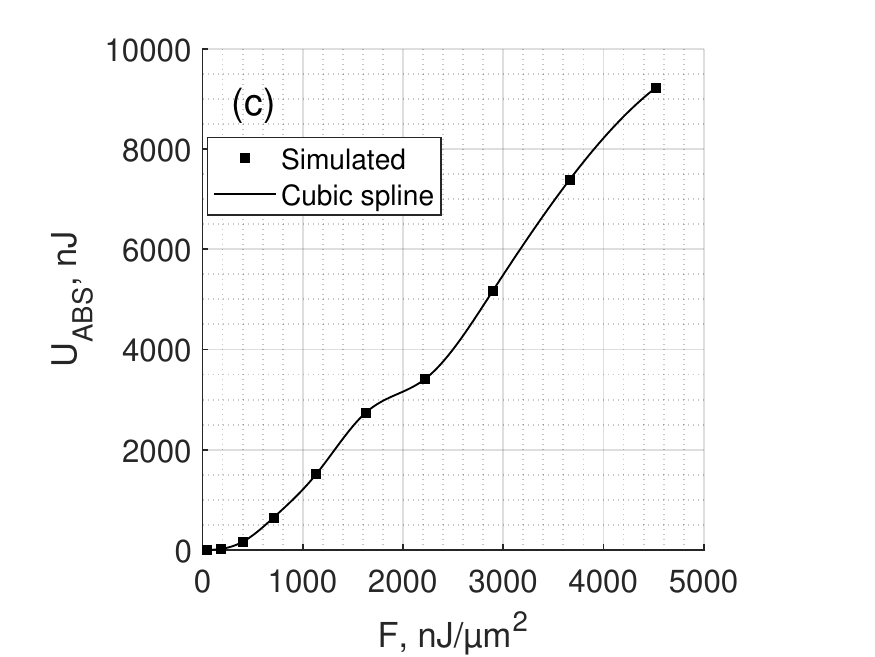}
    \caption{Energy absorbed with respect to fluence of a section representing the silicon ridge waveguide in the case of (a) bulk silicon, (b) silicon-on-insulator, or (c) the waveguide itself. The waveguide section has the same relationship between absorbed energy and fluence as the large-encompassing geometries in (a) and (b), whereas the exposed ridge waveguide in (c) has a significantly enhanced energy absorption at the same fluence. Enhanced energy absorption in (c) is similar in shape to Fig. 
    \ref{fig:4}c, suggesting that the cause is due to field enhancement in the waveguide.  }
    \label{fig:5}
\end{figure}

\section{Conclusion}
In this work, field enhancement of two photon absorption in a silicon ridge waveguide has been reported, achieving above-melting threshold fluence from \( \sim \) 5 \si{\micro J} 50-\si{fs} laser pulse. Modulation of the absorbed field intensity envelope varies with Kerr nonlinear refractive index, which has multiple different reported values. Despite the field intensity itself following a similar trend to the two-photon absorption intensity envelope, uncertainty in the Kerr refractive index affects the absorbed energy via the modulation of the field intensity. Two- to three-dimensional modulation of the imparted laser intensity causes discrepancy in predictions of fluence leading to non-thermal melting, which may challenge predicted laser-ablation in experiment. 

\section{Acknowledgement}
 Special thanks to Ruoheng Zhang, Ying Y. Tsui, Robert Fedosejevs, \& Vien Van for their expert feedback on the simulation parameters and the sharing of the ANSYS Lumerical Optics 2023 license. This research was made possible thanks to funding from the NSERC CGSD and Alliance grants.

 \bibliography{bibliography}
\bibliographystyle{unsrt}

\end{document}


\maketitle

\begin{abstract}
    This document is supplementary material for ``Simulation of silicon ridge waveguide enhanced two-photon absorption from femtosecond pulses." Included are simulation tests of the Taflove \& Goorjian model, demonstration of the two-photon absorption profile for different tuning between the Raman and Kerr terms, and a final derivation of the two-photon absorption model.
\end{abstract}

\section{Simulation test}
\begin{figure}[H]
   \centering
    \includegraphics[width=0.45\linewidth]{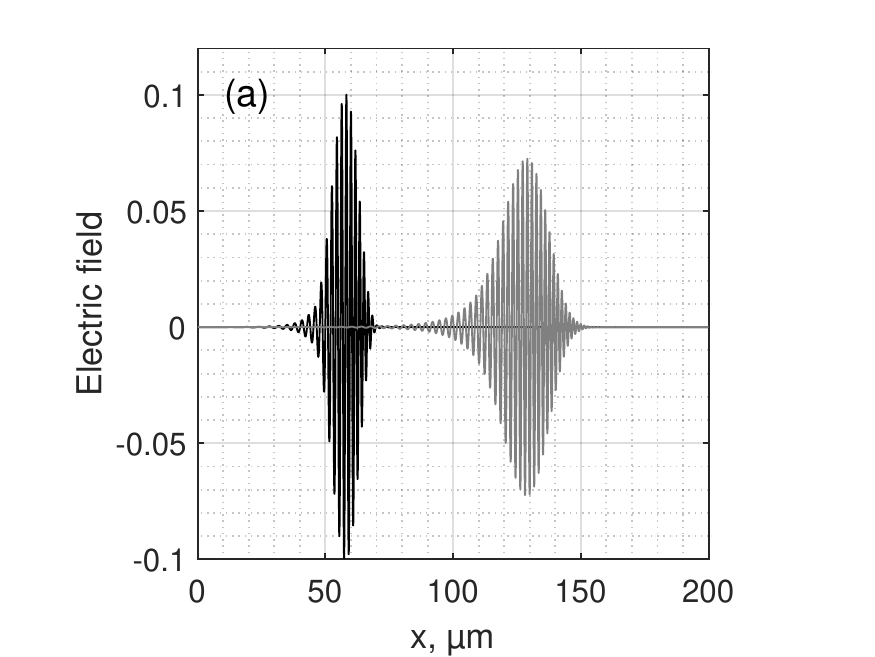}
    \includegraphics[width=0.45\linewidth]{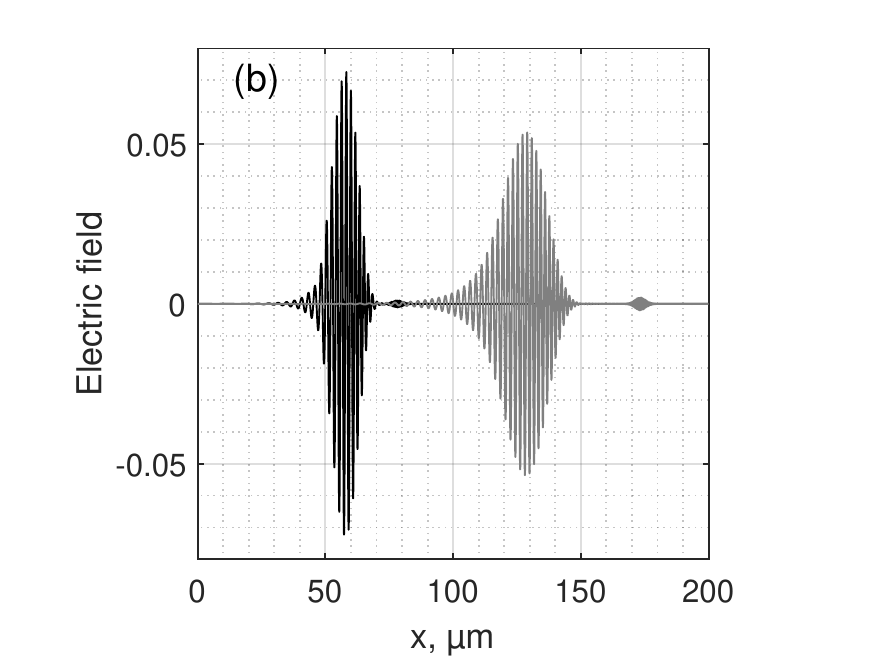}
   \caption{Results for (a) the linear and (b) the nonlinear optical soliton carrier pulse after propagating 55 \si{\micro m} and 126 \si{\micro m} in the Lorentz medium. The solitons appear ahead of the linear carrier pulse, similar to Goorjian \& Taflove's simulations \cite{goorjian1992direct}.}
   \label{fig:1}
\end{figure}

In order to qualitatively confirm the appropriate optical nonlinearity is exhibited by the ANSYS Lumerical Chi3 Raman Kerr material plugin, we repeat simulations performed by Goorjian \& Taflove for temporal evolution of solitons in silica \cite{goorjian1992direct}. Using a custom Drude-Lorentz model as the base material, and given specific nonlinear coefficients in (Tab. \ref{tab:1}), a soliton pulse emerges in front of the incident pulse as they propagate together along a 200 \si{\micro m} waveguide. Frames of the pulse are plotted after it has propagated 55 \si{\micro m} and 126 \si{\micro m}, where both pulses share qualitative agreement with  Ref. \cite{goorjian1992direct} as shown in Fig \ref{fig:1}. We don't explore numerical accuracy of Goorjian and Taflove's simulation, given the unavailability of the data or analytical solution, and instead assume the material model is functional based on its agreement with the analytical solution \cite{li2017nonlinear}. 

\begin{table}[H]
   \scriptsize
   \centering
   \caption{Model properties for Goorjian \& Taflove \cite{goorjian1992direct}.}
   \begin{tabularx}{\textwidth}{|l|l|l|X|}
   \hline
         \textbf{Parameter name} & \textbf{Symbol} & Taflove model & Unit \\\hline\hline
        Relative permittivity & $\epsilon_\infty$ & 2 & - \\\hline
        Lorentz permittivity & $\Delta\epsilon_L=\epsilon_s-\epsilon_\infty$ & 3 & - \\\hline
        Lorentz resonance & $\omega_c$ & $400(10^{12})$ & \si{rad s^{-1}} \\\hline
        Lorentz linewidth & $\delta_c$ & $62.832(10^{9})$ & \si{rad s^{-1}} \\ \hline
        Raman frequency & $\omega_R/2\pi$ & 87.722 &\si{THz}  \\\hline
        Raman spectrum FWHM & $\delta_R/(2\pi)$ & 62.5 &\si{THz}\\\hline
        Susceptibility model parameter & $\chi^{(3)}_0$ & 0.07 & \si{m^2V^{-2}}\\\hline
        Kerr contribution & $\eta$ & 0.7 & - \\ \hline
   \end{tabularx}
   \label{tab:1}
\end{table}

\section{Time-averaged intensity envelope in bulk silicon}
The intensity profile is examined first since it relates directly to the analytical solutions \cite{li2017nonlinear}. Two different values are selected for $\eta$ to compare absorbed energy according to experiments of two-photon absorption (TPA) \cite{bristow2007two}, or time-response of Raman scattering with photoluminescence \cite{aggarwal2011measurement}. Both include the Raman response, but the Raman response is diminished (compared to the experiments of TPA) when $\eta$ is fitted to the Raman time-domain impulse response. Therefore, fitting of the Raman impulse response provides better agreement with the two-photon absorption intensity envelope, as evidence for the generality of the Raman Kerr model in representing TPA.

\begin{figure}[H]
   \centering
    \includegraphics[width=0.45\linewidth]{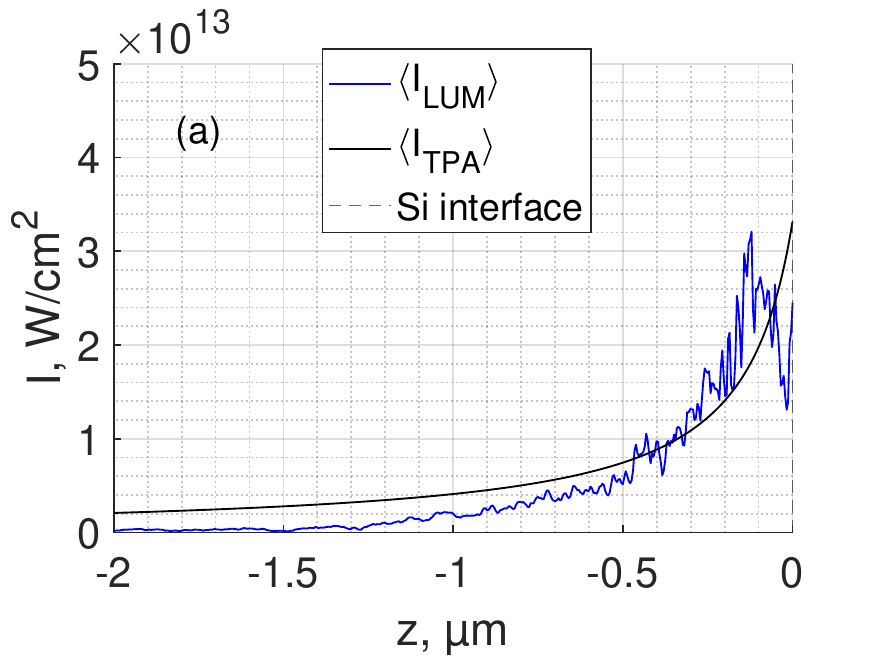}
    \includegraphics[width=0.45\linewidth]{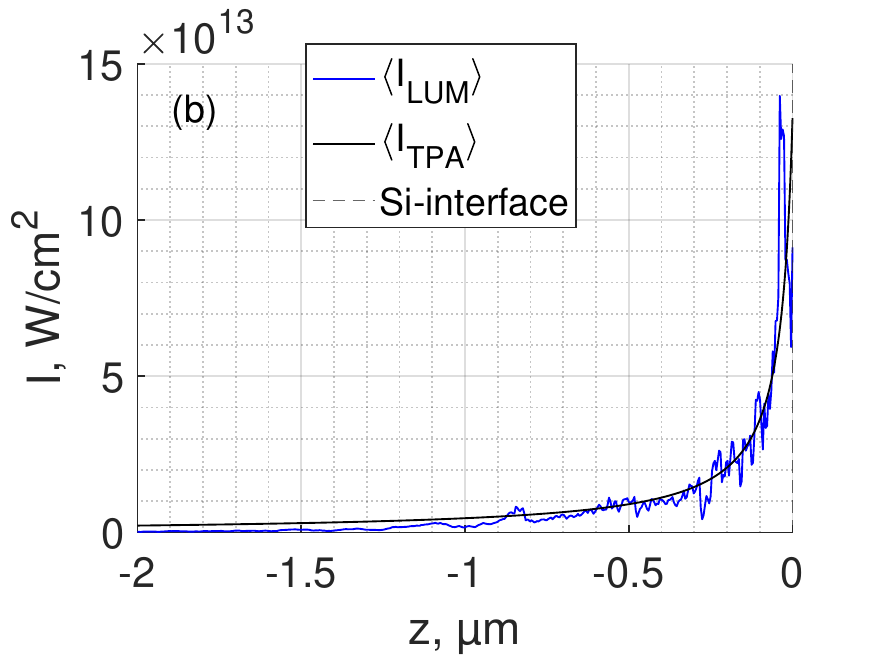}
   \caption{The single-cycle time-average of the intensity near the center wavelength when the pulse has propagated half-way through the interface at $t=t_c+t_d$. The intensity follows the analytical absorption depth profile for each value of $\eta$, but with different field envelope modulations which depends on the refractive index contribution. This figure shows the case for $\eta=0.3575$ based on the impulse response of Aggarwal \emph{et al.}'s measurement \cite{aggarwal2011measurement}, given a pulse energy of (a) 67.08 \si{\micro J} or (b) 94.87 \si{\micro J}. Higher-energy simulations were necessary to demonstrate the increased two-photon absorption. }
   \label{fig:2}
\end{figure}

However, Bristow \emph{et al.} earlier emphasized an enhanced Kerr refractive index measurement significantly greater than the analytically predicted value from Kramers-Kronig relations \cite{bristow2007two}. Bristow \emph{et al.} suggested that the difference was due to a neglect for the Raman contribution. Differences between the two-photon and Raman interpretations has led to distinct measurements of real Kerr refractive index, $n_2$. Two examples of the simulated TPA with Bristow's measured $n_2$ ($\eta=0.001610736$) and the Raman impulse measured $n_2$ ($\eta=0.3575$) are illustrated in Fig. \ref{fig:2}.

\section{Absorbed energy envelope in bulk silicon}
Numerically simulated absorbed time-averaged energy density is expressed as $T\beta\langle\mathrm{I}_\mathrm{LUM}\rangle^2$, whereas the analytical absorbed energy density is expressed as $T\beta\langle I \rangle^2$, where $T$ is the time period of a single cycle (update legends). Note that the analytical solution for $I$ is a time-independent envelope that agrees with the numerical solution. The agreement between the two is dependent on the particular value of $\eta$. In Fig. 1a-c, $\eta=0.001610736$ and there is strong field modulation, whereas in Fig. 1d-e the value for $\eta=0.3575$ is fitted to a Lorentzian Raman spectrum and the agreement of the TPA profile is improved, but non-representative of measured Kerr refractive index \cite{bristow2007two}. This is the present discrepancy between measured and analyzed two-photon absorption at this wavelength.

\begin{figure}[H]
   \centering
    \includegraphics[width=0.32\linewidth]{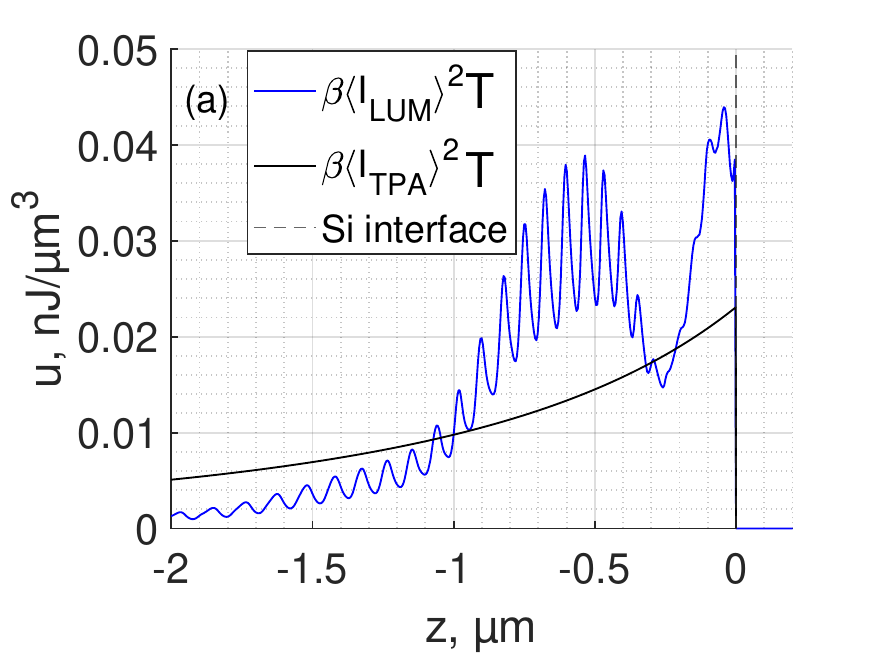}
    \includegraphics[width=0.32\linewidth]{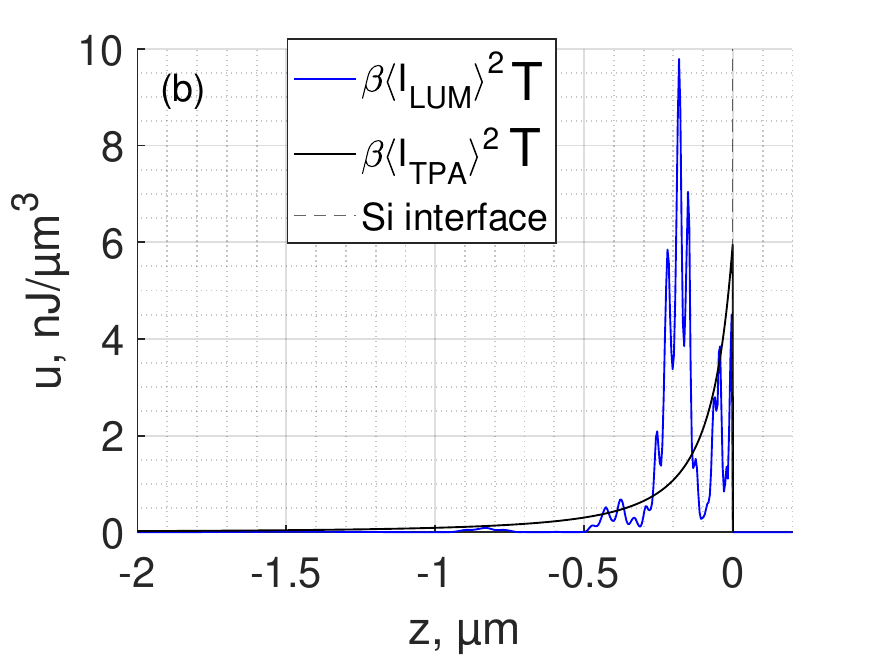}
    \includegraphics[width=0.32\linewidth]{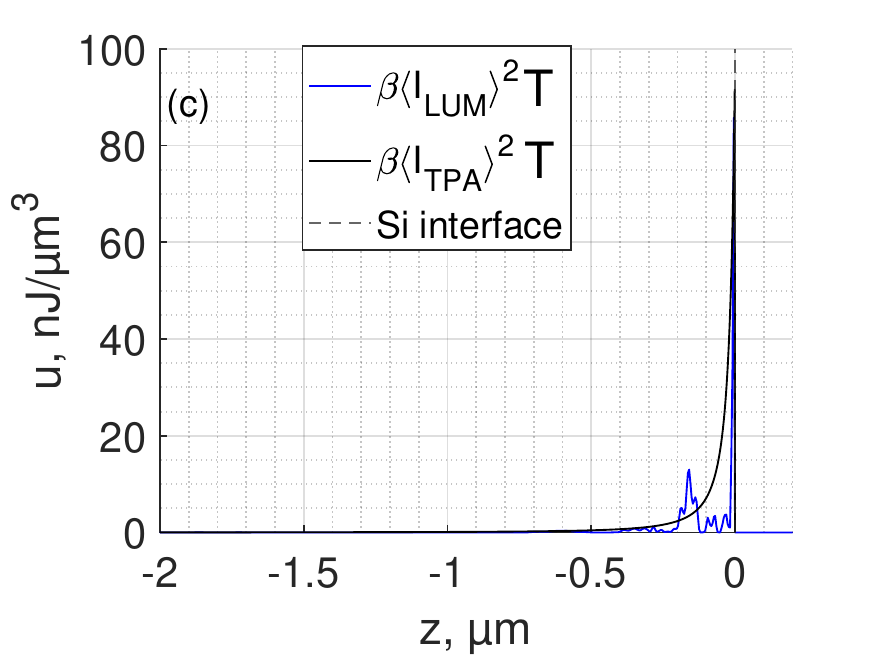} \\
    \includegraphics[width=0.32\linewidth]{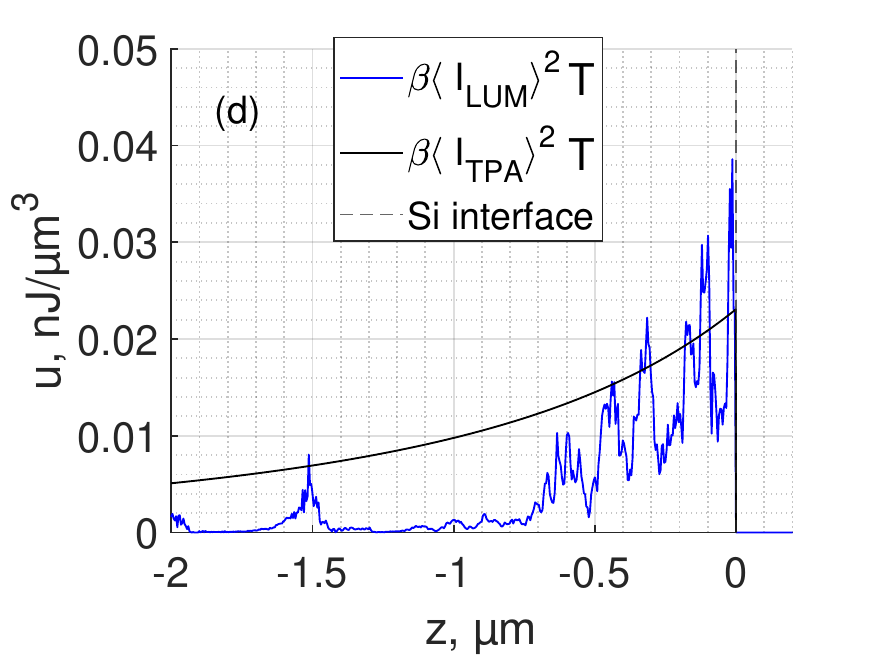}
    \includegraphics[width=0.32\linewidth]{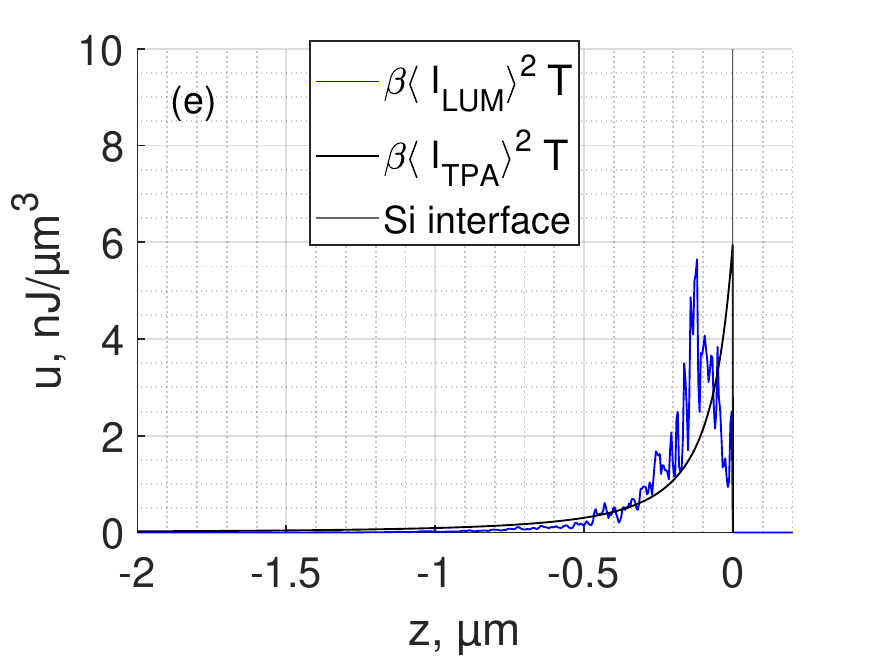}
    \includegraphics[width=0.32\linewidth]{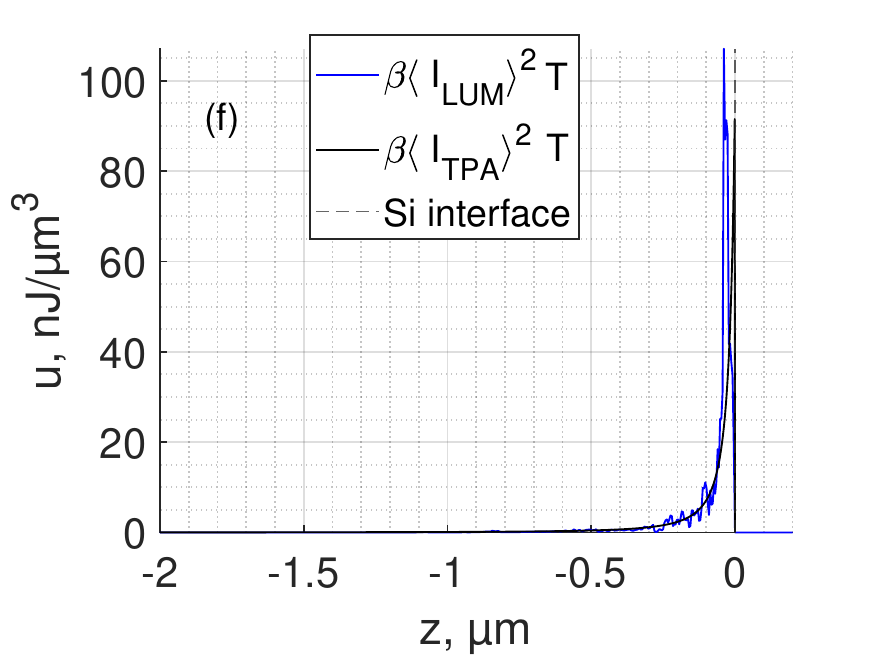}
   \caption{Energy absorption profiles obtained for measured Kerr refractive index, $n_2$, and two-photon absorption coefficient, $\beta$, \cite{bristow2007two} given an energy of (a) 30 \si{\micro  J}, (b) 67.08 \si{\micro J}, and (c) 94.87 \si{\micro J} from a 50-fs pulse. Energy absorption profiles obtained for time-domain Raman response fitting $\eta=0.3575$ \cite{aggarwal2011measurement}, given an energy of (d) 30 \si{\micro J}, (e) 67.08 \si{\micro J}, and (f) 94.87 \si{\micro J} from a 50-fs pulse. Peak energy density is consistent with strong two-photon absorption for either (a-c) or (d-f), but its modulation is slightly different between each case. }
   \label{fig:3}
\end{figure}

\section{Mode confinement}
Linear mode confinement in the silicon structure helps explain the field enhancement. Although the fraction of linearly absorbed energy and two-photon absorbed energy depends on the incident fluence, linear absorption is significant at lower fluence. Multiple linear modes are confined in the structure with the majority of their energy residing in the waveguide and thin film. Transverse magnetic modes are more likely to deposit energy at the surface of the thin film, such that an electric field polarized along the axis of the waveguide may have maximum two-photon absorption. Transverse electric field polarization exhibits more radiating modes in the silicon dioxide layer, which results in additional energy losses compared with a transverse magnetic polarization. 

\begin{figure}[H]
   \centering
    \includegraphics[width=0.9\linewidth]{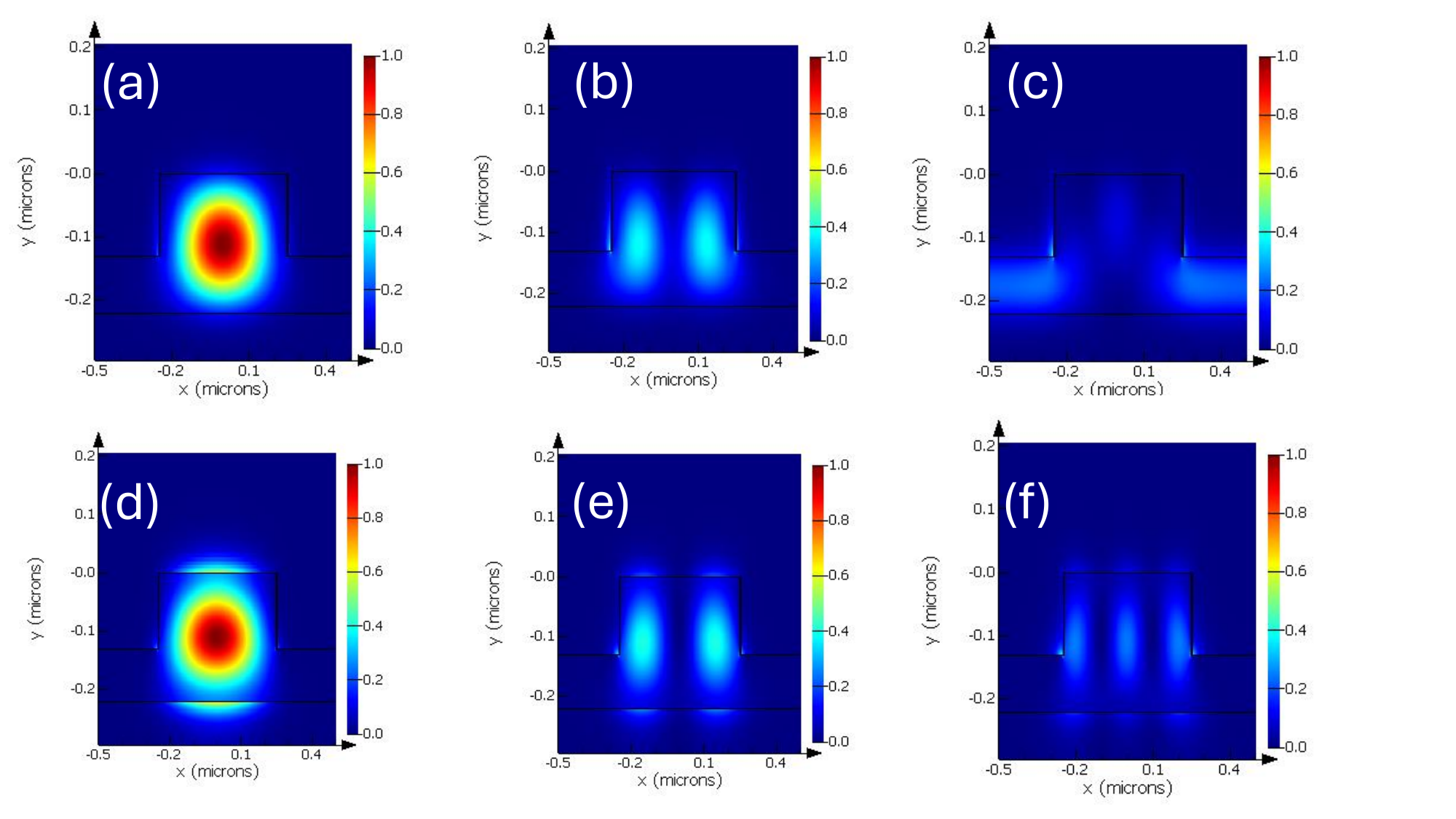}
   \caption{Linear confined modes of the wave-guide for 800 \si{nm} incident wavelength given a quasi (a, b, c) TE polarization; (d, e, f) TM polarization. More TM modes are confined. }
   \label{fig:6}
\end{figure}

\begin{figure}[H]
   \centering    \includegraphics[width=0.9\linewidth]{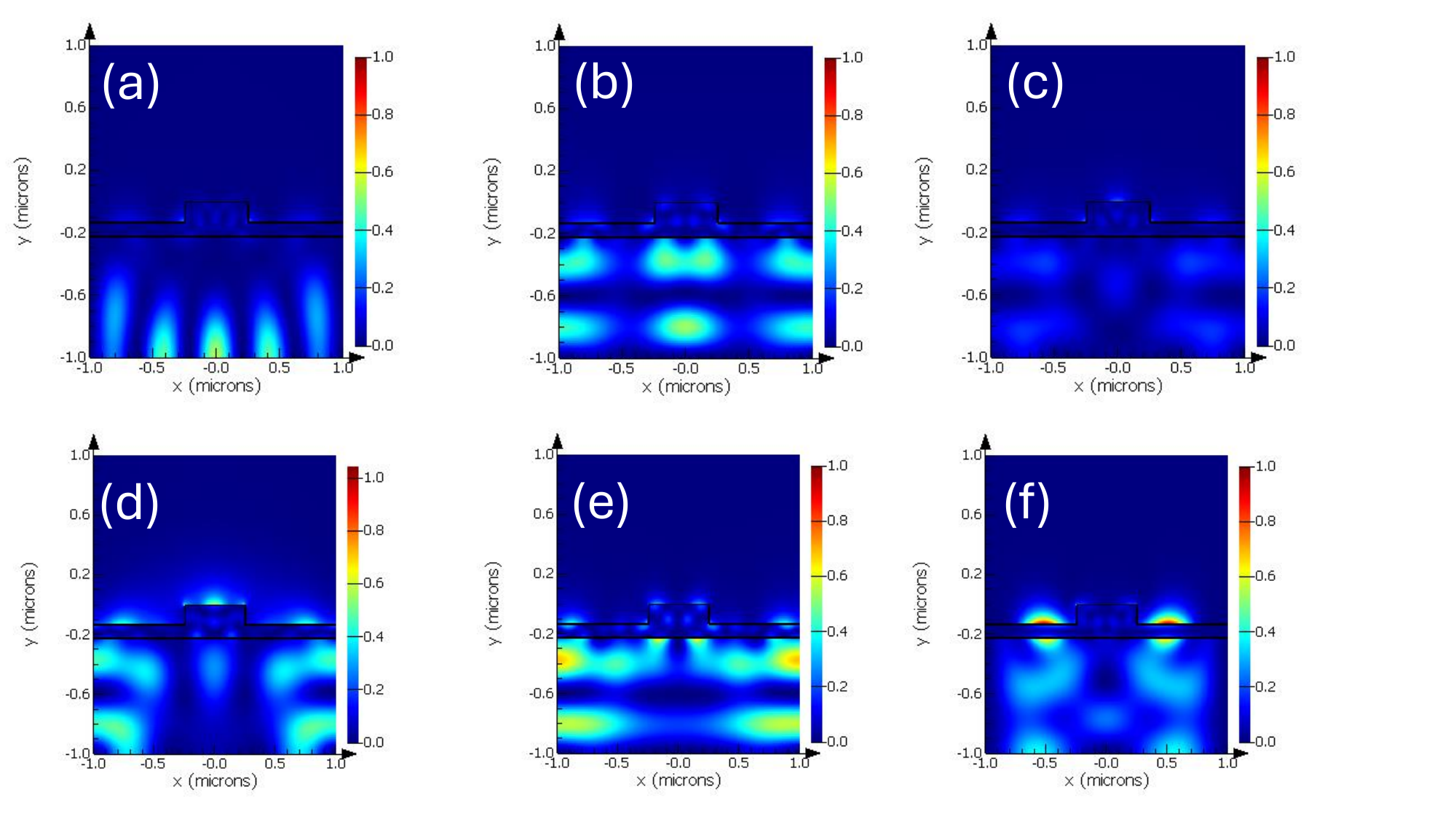}
   \caption{Linear radiating modes of the waveguide for 800 \si{nm} incident wavelength given a quasi (a, b, c) TE polarization; and (d, e, f) TM polarization. More TE modes radiate.}
   \label{fig:7}
\end{figure}

Similar intensity profiles occur from field enhancement with two-photon absorption, except for added filamentation of the modes. Filamentation suggests instability is excited by linear mode confinement, which requires investigation.

\section{Comparison with experiment}
Measurements of the two-photon absorption enhancement have already been recovered by Zhang 2024 \cite{zhang2024precision}. Referring to Fig. 7.7. in Ref. \cite{zhang2024precision}, the blue marker plot datapoints (represented by solid squares) are recovered with permission. The plot represents threshold fluence with respect to number of pulses. Each experiment uses a series of pulses with the same net energy (100 \si{nJ}). Therefore, the total two-photon energy absorbed from a single pulse can be interpreted as the net energy divided by the total number of pulses, $N$, such that

\begin{equation}
    \textrm{U}_\textrm{ABS}=(100\:\textrm{nJ})/N.
\end{equation}

Two-photon absorbed energy is compared with the fluence absorbed exclusively by the silicon material, which is equivalent to the difference in the threshold fluence from its minimum value required for absorption to dominate the light-matter interaction \cite{zhang2024precision}. The fluence absorbed by silicon is less than the threshold fluence measured in experiment, since the latter includes both absorbed and reflected energy.  Assuming dominant two-photon absorption and negligible nucleation and crater formation, such changes in two-photon absorption and geometry of the waveguide may be safely ignored. Therefore, the change in threshold fluence with energy per pulse must be equivalent to the change in absorbed fluence by silicon with respect to two-photon absorbed energy. Consequently, the conversion is
\begin{equation}
    \textrm{F}=\textrm{F}_{\textrm{exp}}-\textrm{F}_{\textrm{exp,min}},
\end{equation}
where the units of $\textrm{F}$ are converted from \si{J/cm^2} to \si{nJ/\micro m^2}.

\bibliography{bibliography}
\bibliographystyle{ieeetr}
\section{Appendix}
First we start with the assumption that the source electric field can be expressed as a sum of its waves at varying frequency (pg. 35-36) \cite{li2017nonlinear}
\begin{equation}\tag{A1}\label{eq:A1}
    \mathbf{E}(\mathbf{r},t)=\sum_j\mathbf{E}(\omega_j)e^{-i(\omega_j t-\mathbf{k}_j\cdot\mathbf{r})}+\sum_j\mathbf{E}^*(\omega_j)e^{i(\omega_j t-\mathbf{k}_j\cdot\mathbf{r})}
\end{equation}
where $\mathbf{r}$ is position, $t$ is time, $\omega_i$ is frequency of index $j$, and $\mathbf{k}_j$ is the corresponding wave-vector.
Assuming that all different frequency waves propagate in the $z-$direction, the $n-$order polarizability in the frequency degenerate case is expressed in terms of a degeneracy factor $D$, such that
\begin{equation}\tag{A2}\label{eq:A2}
    \mathbf{P}^{(n)}(z,\omega)=D\epsilon_0\mathbf{\chi}^{(n)}(\omega;\omega_1,\omega_2,...,\omega_n)\mathbf{E}(z,\omega_1)\mathbf{E}(z,\omega_2)...\mathbf{E}(z,\omega_n),
\end{equation}
where the degeneration factor
\begin{equation}\tag{A3}\label{eq:A3}
    D=\frac{n!}{m!},
\end{equation}
such that $n$ is the order number of the nonlinear polarization, and $m$ is the frequency degenerate number of incident optical waves. In the case of Raman/Kerr nonlinear polarizability, the effect is third order and the frequency degenerate number of optical waves is $2$, such that $n=3, m=2$, and $D=3$. In the case of two-photon absorption polarizability, the effect is third order and the frequency degenerate number of optical waves (photons) is 1, such that $n=3$, $m=1$, and $D=6$. Since both nonlinear polarizability descriptions are of the same order, for a single frequency, they only vary in magnitude based on $D$ and the expression for $\mathbf{\chi}^{(n)}$. In the case of a numerical method, if the two nonlinear polarizabilities are equated, their operation on the electric field and Maxwell's equations must also be equivalent. 

Starting with the quasi-linear approximation in the frequency domain, and assuming plane-wave irradiation, nonlinear polarizabilities affect the spatial variation of electric field as follows (pg. 43-45) \cite{li2017nonlinear}
\begin{equation}\tag{A4}\label{eq:A4}
    \frac{\partial \mathbf{E}(z)}{\partial z} = \frac{i\omega}{2\epsilon_0cn_0}\mathbf{P}_{\mathrm{NL}}(z),
\end{equation}
where $\mathbf{E}$ is the electric field, $z$ is the propagation axis, $i$ is the imaginary unit, $\omega$ is the time-angular frequency, $\epsilon_0$ is the permittivity of free space, $c$ is the speed of light in a vacuum, $n_0$ is the linear refractive index, and $\mathbf{P}_{\mathrm{NL}}(z)$ is the nonlinear polarizability.

Raman/Kerr effects are described using the following nonlinear polarizability

\begin{equation}\tag{A5}\label{eq:A5}
\mathbf{P}_{\mathrm{NL}} = \mathbf{P}^{(3)}(z)=3\epsilon_0\chi^{(3)}|\mathbf{E}(z)|^2\mathbf{E}(z)
\end{equation}
where the factor of 3 arrives from the degeneration factor $D=n!/m!=3!/2!=3$. In contrast, two-photon absorption has two possible nonlinear polarizability options (pg. 184-186) \cite{li2017nonlinear}
\begin{align}\tag{A6}\label{eq:A6}
\mathbf{P}^{(n)}(z,\omega)&=6\epsilon_0\mathbf{\chi}^{(n)}(\omega_1;\omega_2,-\omega_2,\omega_1)\mathbf{E}(z,\omega_2)\mathbf{E}^*(z,\omega_2)\mathbf{E}(z,\omega_1), \\
\mathbf{P}^{(n)}(z,\omega)&=6\epsilon_0\mathbf{\chi}^{(n)}(\omega_2;\omega_1,-\omega_1,\omega_2)\mathbf{E}(z,\omega_1)\mathbf{E}^*(z,\omega_1)\mathbf{E}(z,\omega_2).
\end{align}
There is a symmetric rule of susceptibility where 
\begin{equation}\tag{A7}\label{eq:A7}
    \chi"^{(3)}(\omega_1;\omega_2,-\omega_2,\omega_1)=\chi"^{(3)}(\omega_2;\omega_1,-\omega_1,\omega_2).
\end{equation}
This requires that the two light beams at $\omega_1$ and $\omega_2$ are both absorbed and amplified by the medium simultaneously, with two-photon absorption coefficients for each wave being
\begin{align}
    \beta_1&=\frac{6\omega_1}{\epsilon_0c^2n_1n_2}\chi"^{(3)} \tag{A8a}\label{eq:A8a} \\
    \beta_2&=\frac{6\omega_2}{\epsilon_0c^2n_1n_2}\chi"^{(3)}.\tag{A8b}\label{eq:A8b}
\end{align}
In the case that a single light beam at frequency $\omega$ with intensity $I$ propagates in a medium, it generates a two-photon effect where the two photons have the same frequency $\omega_1=\omega_2=\omega$. This requires that the two light beams also satisfy $I_1=I_2=I$ and $\beta_1=\beta_2=\beta$. Similarly, the nonlinear polarizability for two-photon absorption of this single wave is described by

\begin{equation}
\mathbf{P}_{\mathrm{NL}} = \mathbf{P}^{(3)}(z)=6\epsilon_0\chi^{(3)}|\mathbf{E}(z)|^2\mathbf{E}(z),\tag{A9}\label{eq:A9}
\end{equation}

since the two-photon polarizability is now frequency-independent. This is similar to the case of the off-resonant Raman-Kerr model, except for a factor of 2 which must be accounted for when describing $\chi^{(3)}$ between measurement and simulation. Now the two polarizabilities effect on the electric field can be described by (pg. 43-45) \cite{li2017nonlinear}

\begin{equation}\tag{A10}\label{eq:A10}
    \frac{\partial \mathbf{E}(z)}{\partial z}=D\frac{i\omega}{2cn_0}[\chi'^{(3)}(\omega)+i\chi"^{(3)}(\omega)]|\mathbf{E}(z)|^2\mathbf{E}(z).
\end{equation}
We can substitute the definition for the intensity $I(z)=(1/2)\epsilon_0 c n_0 |\mathbf{E}(z)|^2$
\begin{equation}\tag{A11}\label{eq:A11}
    \frac{\partial \mathbf{E}(z)}{\partial z}=D\frac{ik_0}{2\epsilon_0cn_0^2}[\chi'^{(3)}(\omega)+i\chi"^{(3)}(\omega)]I(z)\mathbf{E}(z),
\end{equation}
or more simply
\begin{equation}\tag{A12}\label{eq:A12}
    \frac{\partial \mathbf{E}(z)}{\partial z}=i\kappa_\mathrm{NL}\mathbf{E}(z),
\end{equation}
where 
\begin{equation}\tag{A13}\label{eq:A13}
    \kappa_\mathrm{NL}(z)=\frac{6k_0I(z)}{\epsilon_0 cn_0^2}\left\{ \chi'^{(3)}(\omega)+i\chi"^{(3)}(\omega)\right\}.
\end{equation}
The role of $\kappa_\mathrm{NL}$ is to cause the attenuation of the electric field as $\mathbf{E}(z)=\mathbf{E}(0)e^{i\kappa_{\mathrm{NL}}(z)z}$. Therefore, the total wavenumber
\begin{align}
    \kappa&=\kappa' + i\kappa"=n+i\frac{\alpha}{2}, \tag{A14a}\label{eq:14a}\\
    \kappa&=k_0n_0+k_0\Delta n + i\left(\frac{\alpha_0}{2}+\frac{\Delta \alpha}{2}\right),\tag{A14b}\label{eq:14b}
\end{align}
has a nonlinear part
\begin{equation}
    \kappa_\mathrm{NL}=k_0\Delta n + i\frac{\Delta \alpha}{2}.\tag{A15}\label{eq:A15}
\end{equation}

We can rearrange the real and imaginary terms to solve for $\Delta n$ and $\Delta \alpha$, assuming $k_0=\omega_0/c$, such that
\begin{align}
     \Delta n &= \frac{6I}{\epsilon_0cn_0^2}\chi'^{(3)}(\omega)\tag{A16a}\label{eq:A16a} \\
    \Delta \alpha &= \frac{12I\omega}{\epsilon_0c^2n_0^2}\chi"^{(3)}(\omega),\tag{A16b}\label{eq:A16b}
\end{align}

and the related experimental values from two-photon absorption $\beta$ and $n_2$, are related by
\begin{align}
    n_2&=\frac{\Delta n}{I}=\frac{6}{\epsilon_0cn_0^2}\chi'^{(3)}(\omega)\tag{A17a}\label{eq:A17a} \\
    \beta&=\frac{\Delta \alpha}{I}=\frac{12\omega}{\epsilon_0c^2n_0^2}\chi"^{(3)}(\omega).\tag{A17b}\label{eq:A17b}
\end{align}
Rearranging the previous two expressions,
\begin{align}
    \chi'^{(3)}(\omega)&=\frac{\epsilon_0cn_0^2}{6}n_2\tag{A18a}\label{eq:A18a} \\
    \chi"^{(3)}(\omega)&=\frac{\epsilon_0c^2n_0^2}{12\omega}\beta. \tag{A18b}\label{eq:A18b}
\end{align}

The final step is to relate these nonlinear susceptibilities to their Raman-Kerr equivalents in the ANSYS Lumerical Chi3 Raman/Kerr plugin \cite{ansysOptics}. One issue may initially appear that this plugin material has a frequency dependence while the two-photon absorption does not. Although this is true, the frequency variation is inconsequential at 800 nm since it is far from the Raman wavelength such that a resonance is unlikely to occur. Instead, the susceptibility maintains the experimental values so long as they are correctly represented by the plugin material model
\begin{equation}\tag{A19}\label{eq:A19}
    \mathbf{P}(t)=\epsilon_0\chi^{(1)}\mathbf{E}(t)+\epsilon_0\eta\chi_0^{(3)}\mathbf{E}^3(t)+\epsilon_0(1-\eta)\mathbf{E}(t)(\chi^{(3)}(t)\star\mathbf{E}^2(t))
\end{equation}
with complex permittivity
\begin{equation}\tag{A20}\label{eq:A20}
\chi^{(3)}=\frac{\chi_0^{(3)}\omega_R^2}{\omega_R^2-2i\omega\delta_R - \omega^2}
\end{equation}
where $\omega_R$ is the Raman frequency, $\delta_R$ is the Raman spectral width, and $\eta$ is a unitless coefficient that weights the proportion of Raman and Kerr nonlinearities. Raman and Kerr phenomena have the same value for $D$. Setting $\eta=0$ will assume that the Raman phenomena is dominant along with the generation of free carriers. Setting $\eta=1$ assumes that the excitation and recombination are instead dominant. This depends on a sensitive experiment, so we leave $\eta=0$. Regardless of $\eta$, setting the real and imaginary components of the Chi3 Raman Kerr plugin material model susceptibility to the real and imaginary susceptibility for two-photon absorption
\begin{align}
\chi'^{(3)}&=\frac{\chi_0^{(3)}\omega_R^2(\omega_R^2-\omega^2)}{(\omega_R^2-\omega^2)^2+4\omega^2\delta_R^2}=\frac{\epsilon_0cn_0^2n_2}{6} \tag{A21a}\label{eq:A21a}\\
\chi"^{(3)}&=\frac{2\chi_0^{(3)}\omega_R^2\delta_R\omega}{(\omega_R^2-\omega^2)^2+4\omega^2\delta_R^2}=\frac{\epsilon_0c^2n_0^2\beta}{12\omega}.\tag{A21b}\label{eq:A21b}
\end{align}
Finally, we use the imaginary term to derive $\chi_0^{(3)}$ which allows full definition of the Chi3 Raman Kerr plugin material in terms of the two-photon absorption coefficients
\begin{equation}\tag{A22}\label{eq:A22}
\chi_0^{(3)}=\frac{\chi"^{(3)}[(\omega_R^2-\omega^2)^2+4\omega^2\delta_R^2]}{2\omega_R^2\delta_R\omega}.
\end{equation}
Substitute $\chi"^{(3)}$ and obtain the form
\begin{equation}\tag{A23}\label{eq:A23}
\chi_0^{(3)}=\frac{\epsilon_0c^2n_0^2\beta}{24\omega^2}\frac{[(\omega_R^2-\omega^2)^2+4\omega^2\delta_R^2]}{\omega_R^2\delta_R}.
\end{equation}

 The fitting parameters can be confirmed by re-substituting them to determine if they return the same Kerr refractive index and two-photon absorption coefficient

\begin{align}
    n_2 &= \frac{6}{\epsilon_0cn_0^2}\frac{\chi_0^{(3)}\omega_R^2(\omega_R^2-\omega^2)}{(\omega_R^2-\omega^2)^2+4\omega_R^2\omega^2}, \tag{26a}\label{eq:A26a}\\
    \beta &= \frac{24\omega^2\chi_0^{(3)}\omega_R^2\delta_R}{\epsilon_0c^2n_0^2\left[(\omega_R^2-\omega^2)^2+4\omega^2\delta_R^2\right]}.\tag{26b}\label{eq:A27b}
\end{align}
It is unlikely that both $n_2$ and $\beta$ satisfy the same as their measured values, given non-Lorentzian Raman spectra. For that reason, the absorption is fitted to $\beta$ and $n_2$ is tuned with $\eta$ to recover the appropriate real susceptibility.